# A Generative Model for Inverse Design of Metamaterials

*Zhaocheng Liu[1], Dayu Zhu[1], Sean P. Rodrigues[1,2], Kyu-Tae Lee[1], and Wenshan Cai[1,2]**

[1] School of Electrical and Computer Engineering, Georgia Institute of Technology, Atlanta, Georgia 30332

[2] School of Materials Science and Engineering, Georgia Institute of Technology, Atlanta, Georgia 30332

* Correspondence should be addressed to W.C. (wcai@gatech.edu)

**Abstract:** The advent of two-dimensional metamaterials in recent years has ushered in a revolutionary means to manipulate the behavior of light on the nanoscale. The effective parameters of these architected materials render unprecedented control over the optical properties of light, thereby eliciting previously unattainable applications in flat lenses, holographic imaging, and emission control among others. The design of such structures, to date, has relied on the expertise of an optical scientist to guide a progression of electromagnetic simulations that iteratively solve Maxwell's equations until a locally optimized solution can be attained. In this work, we identify a solution to circumvent this intuition-guided design by means of a deep learning architecture. When fed an input set of optical spectra, the constructed generative network assimilates a candidate pattern from a user-defined dataset of geometric structures in order to match the input spectra. The generated metamaterial patterns demonstrate high fidelity, yielding equivalent optical spectra at an average accuracy of about 0.9. This approach reveals an opportunity to expedite the discovery and design of metasurfaces for tailored optical responses in a systematic, inverse-design manner.

## 1. Introduction

Near the close of the last century, discoveries in light-matter interactions on the nanoscale unlocked optical phenomena that would help to confine light to sub-wavelength scales, opening a gateway to a new era of optical design. The metamaterial, a member of this family of new nanophotonic devices, is capable of generating periodic dipoles in order to manipulate the behavior of light in a non-classically predicted manner. As such, a well-designed metamaterial can tailor the transmittance and phase delay of electromagnetic waves over any wavelength spectrum [1-5]. The realization of these materials has led to a vast number of applications in perfect absorption [6], super resolution imaging [7,8], beam steering [9,10], and nonlinear optical generation [11,12].

As these nanostructured materials require labor intensive fabrication, an accurate prediction of the optical spectrum and structure of the envisioned metamaterial must be preemptively articulated. However, the complicated physical mechanisms that describe these light-matter interactions at the nanoscale cannot be resolved by generalized theory and as such the prediction of a material's optical properties and approximate structure relies on advanced iterative calculations achieved by Finite-Element Modeling (FEM) or Finite-Difference Time-Domain (FDTD) methods. Moreover, this conventional metasurface design process is innately flawed by human guided error. Not only is the initial design realized based on physical insights and intuitive reasoning, but the finalized geometric and material parameters are ultimately achieved by a means of trial-and-error. The design of such optical systems demands a working knowledge base of optics in order to moderate iterative simulations that scan multi-dimensional parameter spaces. Thanks to rapid developments in artificial intelligence (AI), some scientific problems that classically required human perception or intricate mechanisms have recently been solved by AI [13-17]. Such methods have translated into the field of optics, by employing optimization methods [18-20] and evolutionary algorithms [21] to expedite the design of photonic devices. To accelerate the design process without extensive computation (numerical or analytical) of Maxwell's equations, data driven methods, especially deep neural networks, have been gradually incorporated into the design of microwave and nanophotonics devices [22-25].

With growing interest of new phenomena and applications using metasurfaces, there is an ever-pressing need to develop efficient methods that expedite the discovery and design of novel metasurface structures with custom-defined functionality. As illustrated in Figure 1a, this work aims to leverage deep neural networks to approximate the spectra of a metasurface, and more importantly, to generate metasurface patterns that yield customer-defined spectra at the input. The latter is the long-sought-after goal of inverse optical design, in which a working structure is to be generated directly based on the desired optical responses of the designer. In doing so, the need for extensive parameter scans or trial-and-error procedures is bypassed. However, due to the enormous degrees of freedom in typical metasurface patterns, conventional neural network schema, like backpropagation in trained simulator networks [25], are impotent in the inverse design of metasurfaces. In addition, the trained simulator defines the problem as a deterministic system, such that a single simulator will inevitably lead to a fixed outcome for a given input condition. Meanwhile, multiple solutions may exist for the same target spectrum fed to the simulator at the input, thereby imposing an unnecessary constraint on the diversity of the optimized structures.

To mitigate these challenges, here we adopt a generative adversarial network (GAN) in the network model. A GAN is an unsupervised learning architecture consisting of two networks, namely a generator and a critic, that contesting with each other to create authentic image and video datasets [26-29]. Instead of employing a single simulator to tackle the inverse design problem, we incorporate a GAN to jointly seek the structures for the intended spectra at the input. We have also constructed a geometric dataset that is comprised of images of random shapes. When training the GAN-based network with the geometric data, the resultant patterns from the network will resemble some samples in the geometric dataset. Since constructing such geometric data is largely effortless, given a sufficiently large dataset of the geometries, we will be able to identify and refine proper shapes of the structures in order to replicate the demanded spectra at the input of the networks. Compared to existing practices of optical design with neural networks, which are only able to optimize a few geometrical parameters of a fixed structure represented as vectors (i.e., thickness distribution of layered systems), our network allows for the generation of essentially arbitrary patterns of the unit cell structure, represented as pixelwise images. Moreover, the technique developed here can process multiple input spectra in parallel, which facilitates efficient design and optimization of more than one optical structures simultaneously. Such a need is commonly encountered in metasurfaces research, where gradient meta-structures of varying unit cells are distributed in a two-dimensional space for wavefront shaping applications such as metaholograms, vortex generation, and planar lenses.

**2. Network architecture**

To realize the AI-based optical design described above, we have constructed a network architecture as illustrated Fig. 1b. We divide the network into three parts: a simulator ($S$), a generator ($G$) and a critic ($D$). The primary goal is to train an overfitted generator, which produces metasurface patterns in response to given input spectra $T$ such that the Euclidean distance between the spectra of the generated pattern $T'$ and the input spectra $T$ is minimized. All three networks are convolutional neural networks with delicate differences in detailed structures. The simulator is a pretrained model with fixed weights, taking the generated patterns as input and approximating their transmittances spectra $\hat{T}$ without the use of electromagnetic simulations. It is built to control the accuracy of the optical spectra of the generated structures when training the generator. The generator and the critic together constitute a GAN. The critic of this GAN accepts both the user-defined geometric data and the patterns generated from the generator, and then yields a value $l$ which is essential to compute the distance between the distributions of the two sets of data. By minimizing this distance, the critic network guides the generator to produce patterns that share commons features with the input geometric data. During the training process, we update the weights in the generator by backpropagation from the losses defined by the simulator and the critic. Valid patterns produced by the generator are documented throughout the training process; this occurs whenever the losses of the simulator and critic are sufficiently small. The generated patterns are finally smoothed to binary images as candidates of the metasurface design. Detailed network configurations and training methods are presented in the Supplementary Information.

In essence, the critic learns the distribution of the geometric data and restricts the generator to produce patterns in the image space where the geometric data resides in. There are several benefits of including such a critic into the overall network architecture. First, it excludes a large number of unacceptable patterns that are deemed unrealistic in actual nanofabrication. Second, it allows us to control the overall shape of the generated patterns by feeding similar geometries into the critic. While such forethoughts are not necessary for AI-guided optical design, as demonstrated in the later part of this work, certain geometrical constraints at the input may help to narrow down the potential candidates and thereby expedite the convergence to a solution. Finally, since the production of various categories of geometric data is straightforward and largely effortless, we can modify the distributions of the geometric data to avoid degeneracy. This may occur when multiple metasurface patterns of different topologies possess identical optical spectra to the input spectra, within an acceptable margin of deviation.

As a representative and generalizable case study, we will apply the strategy outlined above to the design of metasurfaces with prescribed spectral behavior under linearly polarized illumination. The general unit cell of the metasurface used in this model is shown in Figure 2a, which has a single layered gold pattern in a square lattice, situated on a glass substrate. Other common parameters include a lattice constant of $w = 340$ nm and a thickness of the gold layer set to $d = 50$ nm. To train the simulator with sufficient data, we carried out 6,500 full wave finite element method (FEM) simulations for metasurfaces with a wide variety of shapes that replaced the metal in the unit cells. The simulation was run over a frequency span from $f = 170$ THz to 600 THz (i.e., 500 nm to 1.8 μm in the wavelength domain), which covers a major portion of the visible and the near-infrared spectral range. These FEM simulations yield the transmittance magnitude spectra $T_{ij}$ of each metasurface under *x*- and *y*-polarized illumination, where *i* and *j* indicates the polarization directions for the incidence and the detection, respectively. Throughout the training, the unit cell structure is represented as a binary image of 64 × 64 pixels, in which 1 stands for gold and −1 for void (air). Each transmission spectrum $T_{ij}(f)$ is represented as a 32-entry vector with equal frequency intervals. The simulator after the training process is able to approximate the transmittance *T* with an average absolute error of less than 0.01 at each frequency point. An example for the effectiveness of the simulator is illustrated in Fig. 2b, where the solid lines are the results from the FEM electromagnetic simulations for the ellipse particle array shown in Fig. 2a. Here, the circles represent the spectral approximation of the same structure obtained by the neural network.

To demonstrate the mechanism and performance of the network, we classify the geometric data into several classes such as circles, arcs, crosses, ellipses, rectangles, etc. A full list of the training data and corresponding samples are made available in the Supplementary Information. Figure 2c illustrates a series of generated patterns during a training process after certain iterations, with and without the critic network. For this example, the critic is fed with a class of geometric data only comprised of crosses. When the critic is on, a cross pattern emerges and gradually adjusts itself to meet the requirement of the input spectrum. In sharp contrast, when the critic network is turned off, the generated pattern is a cluster of random pixels and stays stabilized after a few hundred iterations.

## 3. Results and analysis

Our network architecture is capable of generating metasurface patterns in response to the arbitrarily-input, on-demand optical spectra, whose $T_{ij}$ components and frequency range of interest are defined by the user. At the input of the generator, we specify a set of transmittance spectra $T_{ij}$ from 170 to 600 THz, which corresponds to a wavelength range of 500 nm to 1.8 μm. In the following discussion, we define a set of patterns to be used as a test set and denote this test set as *s*, the spectra of each *s* as *T*. Once these spectra are passed through the network architecture, a pattern is retrieved which we denote as *s'* in correspondence to *s*. In verification, the generated set *s'* are FEM-simulated and defined as *T'*. Unless mentioned otherwise, in the following experiments we set the number of patterns being parallelly searched at each run to be 40 and the total iterations of the training to be 50,000. For each target spectrum, valid patterns may occur at different stages of the training process whenever the losses of the simulator and critic are both reasonably small. On a machine with a single GPU Quadro P5000, it takes approximately 10 minutes to carry out 10,000 iterations of training on average.

As an initial demonstration, to illustrate the overall competence of the proposed framework, we use the spectra, *T*, of randomly selected test samples from each geometric class as the input, and allow the network to seek proper patterns based on these spectra. In this situation, we ensure the existence of solutions by using actual spectra of real patterns as the "target" or input. This constraint will be removed when we perform the inverse design for on-demand spectra, as presented in a later part of the paper. In each test, the critic is fed with 1000 data points (i.e., geometrical shapes) that are randomly generated from the same geometric class. Figure 3a shows representative samples from such experiments for each class of the geometry. The first row depicts the test samples *s*, and the second row shows the corresponding samples *s'* generated by the network. The geometric patterns in each pair of the two rows agree very well, partially because the full spectra input $T_{ij}$ to the generator substantially narrows down the possible solutions. Since the generator does not receive any direct information of the geometry for the input spectra, it may uncover equivalent patterns *s'* that are different from the test structures *s* while yielding the same spectral behavior. Such examples can be found in the cross, sector and arc cases in Fig. 3a, where the discovered patterns *s'* are mirror-flipped counterparts of *s* with the same optical responses under linearly polarized illuminations.

Figure 3b and 3c show the spectra of an ellipse test sample *s* and those of the discovered pattern *s'*, with the corresponding unit cell of the metasurface placed as insets. Comparison between the two sets allows us to conclude that the network has successfully identified the correct structure to replicate the spectra with only minor deviations. We also note that the geometric data fed into the critic network does not necessarily contain the right shape of the resultant solutions. If the type of the right pattern *s* for the required spectra is contained in the geometric data at the input, the nature of the GAN will lead to a decent chance of identifying the structure *s* as a proper candidate; otherwise, the critic will guide the generator to produce patterns with geometric features similar to those of the right geometry [26].

To illustrate the ability of our model to optimize structures of any shape, we further test our network with handwritten digit dataset MNIST as the unit patterns of metasurfaces [30]. We modify patterns from the MNIST dataset with rotation and shifting, and then feed them into the critic during the training. In this experiment, we intentionally exclude digit "5" in the geometric dataset at the input of the critic, and ask the network to generate a pattern that replicates the spectral features of a metasurface with digit "5". Figure 3d and 3e show the spectra of the test sample *s* – a rotated digit "5", and the discovered pattern *s'* – a modified digit "3", respectively. The topologies of the *s* and *s'* differ considerably, but the overall transmittance behaviors of the two samples, especially $T_{xx}$ and $T_{yy}$, possess similar features in terms of both the spectral location and the amplitude. By removing the input structure from the trained data set, the network proves its overarching utility in producing a set of spectra that can competently match that of the input structure.

To quantify the performance of the network, we quantitatively define three types of accuracy as follows. 1) Geometric accuracy: the portion of *s'* that can be recognized as within the same class of the input geometric data; 2) Average accuracy $1-e_{ave}$, where $e_{ave}$ is the average absolute error of the transmittance per frequency point; 3) Minimum accuracy $1-e_{max}$, where $e_{max}$ represents the largest absolute error of the transmittance over all frequencies. Figure 4a displays the three accuracies in our experiments for various classes of geometry. The average and minimum accuracies are calculated based on the correct ones in *s'* in terms of the geometric accuracy. We also note that these accuracies are data-dependent and may vary when the distribution of geometric data changes.

Next, to further exemplify the generality and versatility of the constructed network, we feed into the critic mixed data from all classes of geometry with over 8,000 data points, and the test samples are also randomly selected from all possible geometries. As shown in the last column of Fig. 4a, the accuracy in this situation does not degrade compared to prior studies with only a single geometry input. An example of a discovered structure with mixed geometric input is presented in Fig. 4b and 4c. Just as before, Fig. 4b and 4c indicate the FEM-simulated spectra of the test pattern *s* and the generated pattern *s'* in response to the spectral demand, respectively. Although no specific class of geometry is indicated during the training, the generator is able to reach a pattern, *s'*, that not only geometrically resembles *s*, but more importantly, possesses transmittance spectra, *T'*, nearly identical to that of *T*. In general, if the input geometric data contains more than one topology that satisfies the spectral demand, the network may generate some or all of them in a probabilistic manner. Moreover, by changing the distribution of the geometric data, the user may achieve diverse solutions in response to the same spectral request, thereby mitigating the potential of degeneracy, as described earlier.

As a final example, we demonstrate the efficacy of our approach in the reverse design of a metasurface for user drawn spectral responses, *T*. In practice, the desired spectra at the input are user defined, and the existence of solutions is not guaranteed. This is particularly true when certain constraints are applied to the metasurface design. For instance, in the present study, parameters such as the materials used, the unit cell size, and the thickness of the patterned layer are all predefined in the training data. Nevertheless, when the simulator is sufficiently robust, the network is still able to unearth the best possible pattern that yields spectra *T'* with minimized deviation from the input spectra *T*. To demonstrate this feature, here we design a metasurface with the desired, user-defined spectra behavior shown in the Fig. 4d: i) $T_{xx}$ and $T_{yy}$ are two Gaussian-like resonances with randomly chosen mean $\mu$, variance $\sigma$, and amplitude *a,* and ii) $T_{xy}$ and $T_{yx}$ are zero. The generated pattern along with its spectra *T'* is shown in the Fig. 4e. Although there exists no exact solution to spectral demand described above, the network eventually generates patterns whose spectra share common features with the input spectra including the resonance frequency, the spectral bandwidth, and the transmission magnitude.

## 4. Conclusion and outlook

In this work, we have proposed a generative, deep, network model which efficiently discovers and optimizes unit cell patterns of metasurfaces in response to user-defined, on-demand spectra at the input. Our model addresses the pursuit of reverse design technology for photonic structures, and helps to relieve computational and specialist resources in traditional metasurface design from iterative simulations and parameter sweeping, up to generalizing the process for users lacking a solid knowledge base in optics. The model developed here is mostly based on the unsupervised learning, which guarantees efficient generation of structural patterns independent of human experience. This feature is crucial in the investigation of new structures and novel phenomena in optics and beyond. In addition, as our model can process multiple input spectra without the loss of efficiency, the workload for complicated problems that require multiple metasurfaces or gradient structural distributions will be significantly reduced. The performance of the model can be further improved by applying certain refinements, such as the use of a more sophisticated network configuration and the introduction of loss functions with more practical physical meanings.

The methodology we have developed is readily extended to the pursuit of desired complex values of the reflection and transmission coefficients, which is essential to the applications like meta-lenses and meta-holograms where both the amplitude and the phase of the light waves matter. In the current work, we restrict the unit pattern of the metasurface as a single metallic particle with continuous topology. By revising the geometric dataset and refining the simulator, unit cells with multiple particles and complicated structures can be designed and optimized as well. The increased arbitrariness of topology can significantly improve the spectral accuracy of the outcome from our model. Moreover, if sufficient data for the training of the simulator is available, greater degrees of freedom in the design can be made by slightly modifying the network architecture to include modifications of the lattice constant or optimization of the pattern thickness. The developed framework can be adapted to many other applications in optics and materials science, including photonic crystals, 3D metamaterials, imaging systems, phase transitions, etc. We envision broad and growing utilization of the deep learning technology in the physics realm, so that scientists and engineers will be largely relieved from tedious processes of trial and error, and instead focus more on truly creative thoughts yet to be broached by the machine.

**Figures**

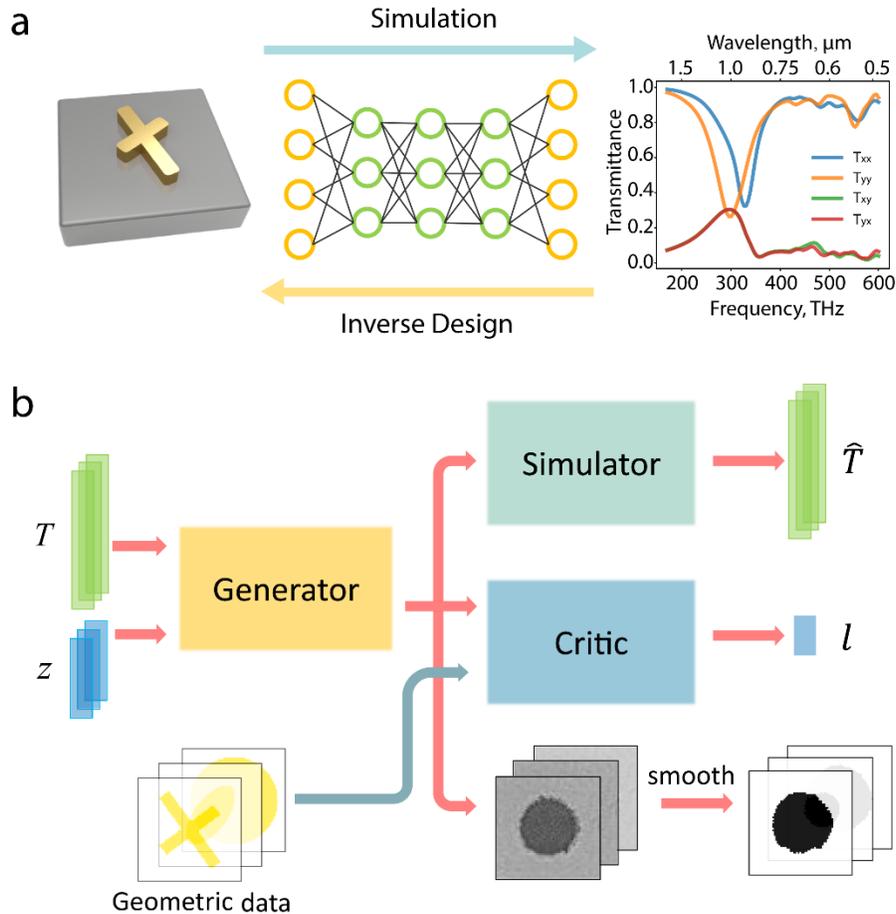

**Figure1 | Transitioning metasurface design from conventional trial-and-error approaches to neural network mediated inverse design**. (a) Both, simulation and inverse design, enact structure-property relationships to generate an optical spectrum from a metasurface and vice-versa. In this work, both processes will be replaced by deep neural networks. (b) Architecture of the proposed network for AI-based optical design. Three networks, the generator (*G*), the simulator (*S*), and the critic (*D*) constitute the complete architecture. The generator accepts the spectra *T* and noise *z* and produces possible patterns. The simulator is a pretrained network that approximates the transmittance spectrum $\hat{T}$ for a given pattern at its input, and the critic evaluates the distance of the distributions between the geometric data and the patterns from the generator. While training the generator, the produced patterns vary according to the feedback obtained from *S* and *D*. Valid patterns are documented during the training process, and are smoothed to qualify as candidate structures.

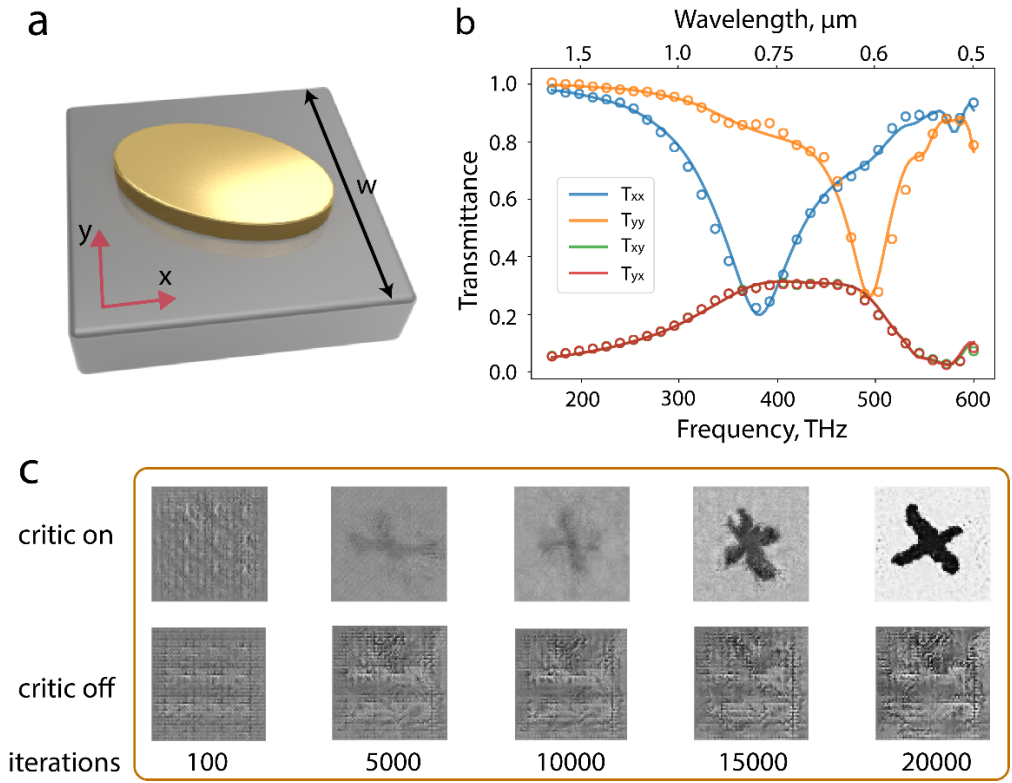

**Figure 2 | Effectiveness of the engineered simulator and critic networks.** (a) The unit cell of the metasurface used in our case study of the inverse design problem. The pattern is allowed to vary for a desired spectral response, with the following structural and material constraints enforced: gold for the pattern and glass as the substrate; unit cell size $w$ = 340 nm; thickness of gold $d$ = 50 nm. (b) Transmission spectra of a representative structure shown in (a), obtained by FEM electromagnetic simulation (solid lines) and by the simulator network (circles), respectively. (c) Generated patterns during the training process after certain iterations, with and without the critic network. Only geometric data categorized as a cross are fed into the critic in this example.

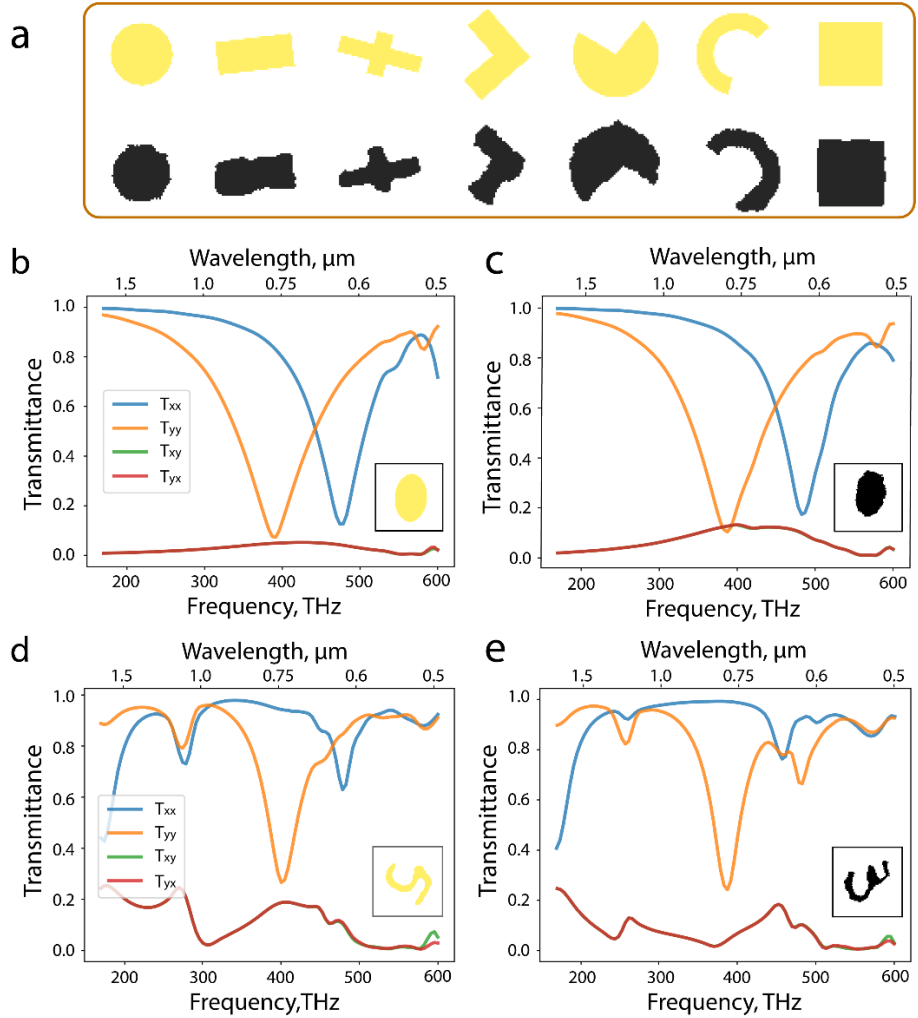

**Figure 3 | Generating patterns with a predesigned class of geometric data.** (a) Test patterns *s* are depicted in the top row and the corresponding generated patterns *s'* are listed in the bottom row. Each shape provides a sample of the different classes of geometric data input to the critic network. (b) Transmittance spectra, *T'*, of a test pattern *s*, to be fed to the network. (c) FEM simulated transmittance of the retrieved pattern *s'*, from the neural network based on the input as in (b). The unit cells of *s* and *s'* are shown in the lower right corner of each figure. This result is achieved when the critic only receives geometric data of the elliptical class. (d) and (e) An example of results with a modified MNIST handwritten digital dataset as the input, geometric data. Note that in this experiment, we intentionally excluded digit "5" in the input geometric data.

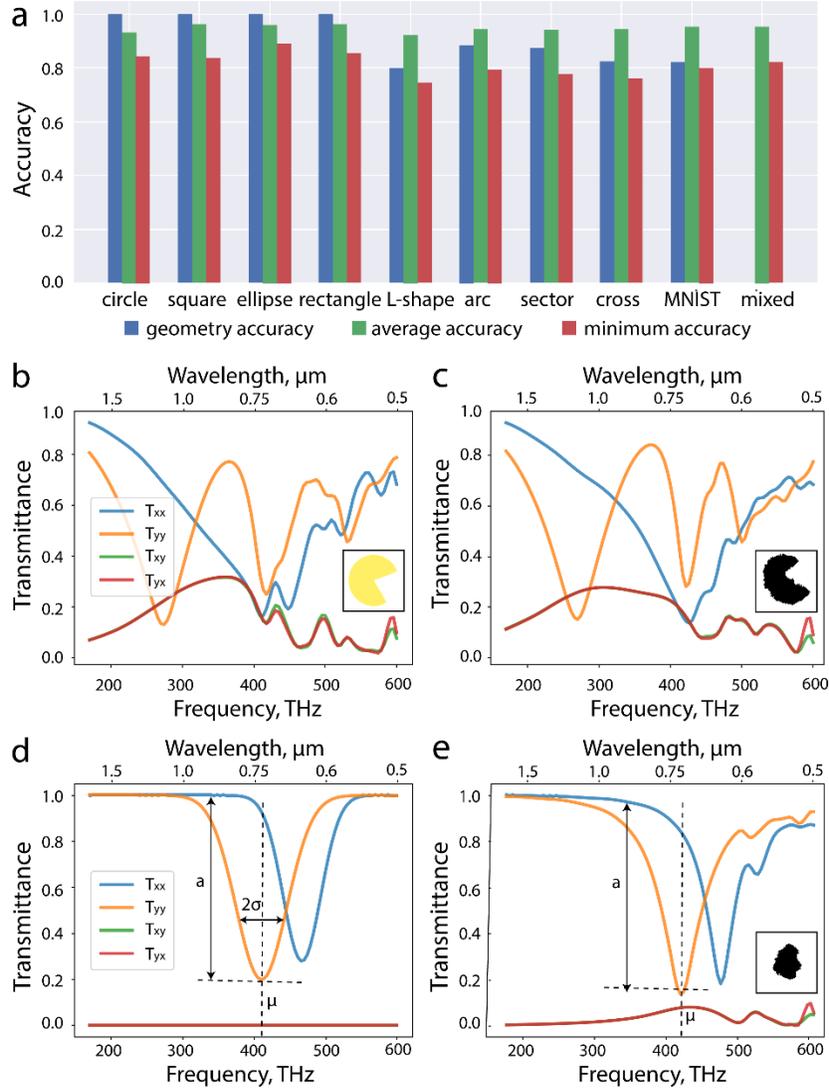

**Figure 4 | Statistical accuracy of the model and examples of generated patterns with mixed classes of geometric data**. (a) Geometric accuracy, average accuracy, and minimum accuracy of experiments with the different classes of geometric data. (b) and (c) Example of results with a mixture of different classes of geometric data used at the input. Transmittance spectra of the test structure *s* and the generated pattern *s'* are shown in the (b) and (c), respectively, with the unit cell depicted as the inset in each figure. (d) and (e) Example of Inverse design of metasurfaces with human-defined spectra. (d) Desired transmittance spectra as the input to the generator, where $T_{xx}$ and $T_{yy}$ are two randomly generated Gaussian-like responses with parameters $a$, $\mu$, and $\sigma$, while $T_{xy}$ and $T_{yx}$ are 0 throughout the frequency range of interest. (e) The resultant unit cell generated by our AI model to fit the target spectra, along with the FEM simulated transmittance spectra of this generated metasurface.